\documentclass[pra,aps,epsfig,graphics,twocolumn,twoside]{revtex4}
\usepackage[breaklinks]{hyperref}
\usepackage[dvips]{graphicx}
\usepackage{url}
\usepackage{breakurl}

\usepackage[T1]{fontenc}
\usepackage[latin2]{inputenc}

\usepackage{amsmath,amssymb}
\usepackage{amsthm}
\usepackage{latexsym}
\usepackage{bbm}
\usepackage{soul}
\usepackage{graphicx}
\usepackage{epsfig}
\usepackage{color}

\def\dt{{\rm d}\,}

\def\tblue{\textcolor{black}}

\def\duzomniejsze{<\kern-.7mm<}
\def\duzowieksze{>\kern-.7mm>}

\def\textbf#1{{\bf #1}}
\def\beq{\begin{equation}}
\def\eeq{\end{equation}}
\def\be{\begin{equation}}
\def\ee{\end{equation}}
\def\ben{\begin{eqnarray}}
\def\een{\end{eqnarray}}
\def\beqa{\begin{eqnarray}}
\def\eeqa{\end{eqnarray}}
\def\eea{\end{array}}
\def\bea{\begin{array}}
\newcommand{\bei}{\begin{itemize}}
\newcommand{\eei}{\end{itemize}}
\newcommand{\bee}{\begin{enumerate}}
\newcommand{\eee}{\end{enumerate}}

\def\1{\openone}

\def\>{\rangle}
\def\<{\langle}
\def\blacksquare{\vrule height 4pt width 3pt depth2pt}

\def\ot{\otimes}

\def\dt#1{{{\kern -.0mm\rm d}}#1\,}
\def\squareforqed{\hbox{\rlap{$\sqcap$}$\sqcup$}}
\def\qed{\ifmmode\squareforqed\else{\unskip\nobreak\hfil
\penalty50\hskip1em\null\nobreak\hfil\squareforqed
\parfillskip=0pt\finalhyphendemerits=0\endgraf}\fi}

\newtheorem{lemma}{Lemma}

\newtheorem{theorem}[lemma]{Theorem}
\newtheorem{proposition}[lemma]{Proposition}
\newtheorem{definition}[lemma]{Definition}

\newtheorem{fact}[lemma]{Fact}

\def\bep{\begin{proposition}}
\def\eep{\end{proposition}}
\def\bel{\begin{lemma}}
\def\eel{\end{lemma}}

\def\bet{\begin{theorem}}
\def\eet{\end{theorem}}
\def\bed{\begin{definition}}
\def\eed{\end{definition}}
\def\bef{\begin{fact}}
\def\eef{\end{fact}}

\begin{document}

\title{Long-distance quantum communication over noisy networks without long-time quantum memory}

\author{Pawe\l\  Mazurek$^{1}$, Andrzej Grudka$^{2}$, Micha\l\   Horodecki$^{1}$, Pawe\l\  Horodecki$^{3}$,\\ Justyna \L{}odyga$^{2}$, 
\L ukasz Pankowski$^{1}$ and Anna Przysi\k{e}\.zna$^{1}$}

\affiliation{
$^1$Institute for Theoretical Physics and Astrophysics,
University of Gda{\'n}sk, 80-952 Gda{\'n}sk, Poland\\
$^2$Faculty of Physics, Adam Mickiewicz University, 61-614 Pozna\'{n}, Poland\\
$^3$Faculty of Technical Physics and Applied Mathematics,
 Gda{\'n}sk University of Technology, 80-952 Gda{\'n}sk, Poland
}


\begin{abstract}
The problem of  sharing entanglement over large distances is crucial for implementations 
of quantum cryptography. A possible scheme for long-distance entanglement sharing and quantum communication 
exploits networks whose nodes share Einstein-Podolsky-Rosen (EPR) pairs. In Perseguers {\it et al.}  [Phys. Rev. A \textbf{78}, 062324 (2008)] the authors
 put forward an important isomorphism between storing quantum information in a dimension $D$ and  transmission of quantum  information in a $D+1$-dimensional network. We show that it is possible to obtain long-distance entanglement in a noisy two-dimensional (2D) network, even when taking into account that encoding and decoding of a state is exposed to an error. For 3D networks we propose a simple encoding and decoding scheme based solely on syndrome measurements on 2D Kitaev topological quantum memory.
Our procedure constitutes an alternative scheme of {\it state injection} that can be used for universal quantum computation 
on 2D Kitaev code. It is shown that the encoding scheme is equivalent to teleporting the state, from a specific node into a whole two-dimensional network, through some virtual EPR pair existing within the rest of network qubits. We present an analytic lower bound on fidelity of the encoding and decoding procedure, using as our main tool a modified metric on space-time lattice, deviating from a taxicab metric at the first and the last time slices.

\end{abstract}

\maketitle

Suppose we have a network of laboratories with some fixed distance between neighboring ones, 
and one of them wants to  establish quantum communication with another one. We assume that neighboring labs can directly exchange quantum communication with some small, fixed error. 
This can be used e.g. to share some noisy Einstein-Podolsky-Rosen (EPR) pairs between the neighboring labs.  We also assume that all operations performed within each lab may be faulty with some fixed, small probability.
If two distant labs can achieve quantum communication with the help of all the labs in the network, then they can exploit it to share cryptographic key that will be known only to these two labs.
This scenario was put forward in \cite{perseguers-2008-78}, 
and is an alternative to quantum repeaters \cite{repeatersPRL,DuanLCZ2001-long}. 
It is also closely related to entanglement percolation \cite{AcinCiracL-perc}. In \cite{perseguers-2008-78} the question was posed whether for a 2-dimensional network,
in principle, one can perform quantum communication over an arbitrary 
distance, provided that one can execute gates between the adjacent nodes (i.e. local gates), and the size of the system in each node 
is constant (i.e. it does not depend on the distance), so that the nodes do not need quantum memory.
The answer was affirmative. Namely, the authors represented nearest neighbour quantum computation on a line as a teleportation process on quantum 2-dimensional square networks, where entangled pairs are shared between adjacent nodes (i.e. between those that are separated by a size of an elementary cell $a$ of the network). 1-dimensional system for quantum computation is formed by all nodes belonging to a chosen line forming a diagonal of elementary cells, so that nodes on this line are separated by a distance $\sqrt{2}a$. Exploiting entanglement shared between adjacent nodes, the state of every node belonging to 1-dimensional system can be teleported (with certain fidelity) (i) to the node on the right, and from there - (ii) to the node above. In this way the state of the whole 1-dimensional system is teleported to a line parallel to the original one. By associating time with every parallel line one can model a storage of a logical state of 1-dimensional system. This can be expanded to model nearest neighbour quantum computation when one performs qubit unitary operators on nodes after the stage (i) and modifies the teleporting scheme so that qubits from nearest neighbour nodes can be teleported to the same node at stage (i) to perform double qubit gate. In this sense a computation problem on a logical state of 1-dimensional system is equivalent to its teleportation in 2-dimensional network. In \cite{StephensFH2007-local-FT} a scheme for universal fault-tolerant quantum computing in 1 dimension was designed. The fundamental problem with the above scheme from the communication perspective is that it is based
on logical qubits as input states while long distance communication requires good 
transmission between physical nodes.  \\

In this paper we overcome this problem, and 
present the first complete proof of possibility of long distance quantum communication in 2D network with no long-time quantum memory: While calculating the fidelity $F$ we took into account not only the fidelity of a qubit storing $F_{s}$ (determined by the applied error-correction scheme, whose error probability goes down exponentially 
with the size of code), but also fidelities $F_{enc}$, $F_{dec}$ of encoding/decoding a physical qubit in {\it unknown state}
into/from a code. This constitutes our first result stated in Proposition $\ref{prop1}$. As the second one we present a new scheme for encoding physical qubit into a 2D Kitaev topological code \cite{KitajewPreskill-FTmem, Bravyi} considerably simpler than already existing scheme of \cite{perseguers-2008-78} (cf. earlier works \cite{raussendorf20062242,raussendorf2007} with numerical analysis of encoding).
 Our scheme, with analytical bound on fidelity expressed by Proposition $\ref{prop6}$, enables quantum communication in 3D network, as well as can be applied to universal quantum computing architecture based on 2D Kitaev code.\\

In the following we will first present the mentioned results of encoding/ decoding physical qubits in 1D and 2D quantum codes. 
Section \ref{sec:1} is devoted to 1D concatenated code, while encoding/decoding into/from a 2D  Kitaev topological code is considered in Section \ref{sec:2}, under the assumption of noise acting only on data qubits (non-fault tolerant scenario).  
The first result proves that quantum communication between two distant nodes 
of  2D network is possible, while the second provides a simple scheme of communication over 3D network. We proceed with fault-tolerant encoding algorithm in Section \ref{sec:4}, where we derive the lower bound for an associated threshold value for communication in 3D by using a metric on a space-time lattice that takes into account effects of encoding and decoding. In Section \ref{sec:5} we discuss the relation of our results to entanglement percolation \cite{AcinCiracL-perc}, which is closely related to 
quantum communication over networks. We conclude in Section \ref{sec:6}.

\section{Encoding physical qubit into 1D concatenated code.}\label{sec:1}
We start with a physical qubit 
$a|0\>+b|1\>$ and encode it by use of fault tolerant scheme in 1D based on concatenation \cite{AharonovBO1996-FT}. We estimate the fidelity of reaching 
a given level of concatenation $r$ and obtain a bound on the fidelity, that does not depend on $r$, which is concluded in Proposition \ref{prop:1}. 

We can assume that fidelity of decoding is no worse than fidelity of encoding (this happens
for the fully unitary fault-tolerant scheme as e.g. in \cite{AharonovBO1996-FT},
and if we do not need to perform correction, as is the case of cryptographic applications).

Let $v$ be the volume of the physical circuit (i.e. the number of gates, including 
identity gates) which encodes into a logical qubit in a first concatenation level. 
The probability of success in this encoding stage is then larger than 
$(1-p)^v$ where $p$ is the probability of an error per gate. Indeed, if every element of our circuit 
works -- which happens with the probability ($1-p$) -- then the output is correct.  In the next encoding stage, the effective probability 
of an error per logical gate is $p_1\leq cp_0^2$, where $c$ is the number of different pairs of circuit gates,
i.e. $c={v \choose 2}$, and $p_0\equiv p$. 
The probability of successful encoding into a second level of concatenation is no smaller than 
$(1-p_1)^v$.
The probability $p_s^{(r)}$ that we pass successfully $r+1$ stages is a product of 
such probabilities in each stage. Hence, we have 
\be
p_s^{(r)}=(1-p_0)^v(1-p_1)^v \ldots  (1-p_r)^v \geq \prod_{k=0}^r\biggl(1-\frac1c(cp)^{2^k}\biggr)^v.
\ee
Now, we proceed to estimate this from below. Using the notation $\alpha=1/c$, $\beta=cp$ we obtain 
\be
\label{sum_geq_int}
\frac1v \ln(p_s^{(r)})\geq\sum_{k=0}^r\ln(1-\alpha \beta^{2^k})\geq 
\int^{r}_{-1}\ln(1-\alpha \beta^{2^x})\dt x.
\ee
Extending the limit to infinity and changing the variables, we obtain
\be
\frac1v \ln(p_s^{(r)})\geq -\frac{1}{\ln2}\int^{\sqrt{\beta}}_0\frac{\ln(1-\alpha z)}{z\ln z}\dt z.
\ee
Since we assume that $cp<1$ (otherwise the concatenation scheme 
would be useless), the integrated function $f(z)=\frac{\ln(1-\alpha z)}{z\ln z}$ is monotonically increasing,
so we can estimate the integral by $\beta f(\beta)$, obtaining 
\be
\frac1v \ln(p_s^{(r)})\geq -\frac{\ln(1-\sqrt{p/c})}{\ln2 \ln (\sqrt{cp})}\geq \frac{\sqrt{p/c}}{\ln2 \ln (\sqrt{cp})},
\ee
where we used $\ln x\leq x-1$ for all $x \geq 0$. If $cp\leq 1/e$
(which is only slightly stronger than the fault-tolerant threshold assumption 
$cp<1$), we finally obtain: 
\be
\label{prop}
\frac1v \ln(p_s^{(r)}) \geq - 2\sqrt{p},
\ee
which proves:

\begin{proposition}\label{prop1}
If an error rate $p$ satisfies  $p\leq p_{th}/3$, where $p_{th}$ is a threshold 
for an error rate in fault-tolerant architecture based on concatenated codes, 
then the fidelity of encoding procedure satisfies 
\be
F_{0}\geq e^{-2v\sqrt{p}},
\ee
where $v$ is the volume of an encoding circuit on a physical level.
\label{prop:1}
\end{proposition}

It is worth mentioning that the proposition can also be formulated in a bit different way. Instead of (\ref{prop}) we can put
\be
F_{0}\geq (1-p)^v e^{-pv},
\ee
which is a slightly better constraint, but not that nice in a form. This estimation can be obtained in complete analogy to the previous one if the sum in (\ref{sum_geq_int}) is lower bounded as follows: 
\be
\sum_{k=0}^r\ln(1-\alpha \beta^{2^k})\geq 
\ln{(1-\alpha \beta)}+\int^{r}_{0}\ln(1-\alpha \beta^{2^x})\dt x.
\ee

 Therefore, the fidelity $F$ of encoding/storage/decoding of a qubit can be estimated as $F\geq F_s F_{0}^{2}$, where $F_0\geq \exp(-2v\sqrt{p})$. For depolarizing noise $\Lambda(\rho)=p(1-p)X\rho X+p^{2}Y\rho Y+p(1-p)Z\rho Z+(1-p)^2\rho$,  acting on every qubit in one time step of storage, it can be shown that $F_{s}\geq 1- (e-1)T(\frac{2p}{2p_{th}})^{(t+1)^{k}}$, with number of time steps T, threshold probability $p_{th}=\frac{1}{2}\bigg(e{w \choose t+1}\bigg)^{-1/t}$, $k$ enumerating a concatenation level, and $w$, $t $ being the volume of physical circuit implementing a gate and the code distance, for $k=1$, respectively \cite{Aliferis2007, LidarBrun2013}. Therefore, we have $F_{s}>1-c_{1}T\exp(c_{0}V\ln{\frac{p}{p_{th}}})$, with constant $c_{1}=1-e$, and $c_{0}=(\frac{t+1}{N})^k$  such that $V=N^{k}$ is the number of code qubits for a given $k$. We take $p\leq\frac{p_{th}}{3}$ in order to be consistent with the derivation of $F_{0}$, and obtain $F_{s}>1-c_{1}T\exp(-c_{2}V)$, with $c_{2}=\ln{3}(\frac{t+1}{N})^k$. As a result, we can estimate $F>\exp(-2v\sqrt{p})\big(1-2c_{1}T\exp(-c_{2}V)\big)$.

As we will discuss in Sec. \ref{sec:5}, the problem of storing a qubit in unknown state in 1D by means of local gates is equivalent to transmission 
of a qubit in 2D network (as shown in \cite{perseguers-2008-78}), therefore our proposition implies that long distance quantum communication is possible in 2D noisy network. 

A disadvantage of the fault tolerant schemes based on concatenated codes 
is that they are usually quite complex. Moreover, originally, they involve non-local 
gates (i.e. ones which do not connect adjacent qubits). Since swap operators do not propagate errors,
the fault-tolerant scheme of e.g. \cite{AharonovBO1996-FT}  can be built of solely local gates, by 
accompanying any non-local gate with series of swap operators,  
but this further increases complexity of the scheme (see \cite{Gottesman2000,Svore2005,StephensFH2007-local-FT} for various more smart schemes with local gates).
 
\section{Non-fault tolerant encoding/decoding schemes for 2D Kitaev code.}\label{sec:2}
\subsection{Encoding/decoding algorithm}
\begin{figure}
  \centering
  \includegraphics[scale=0.25]{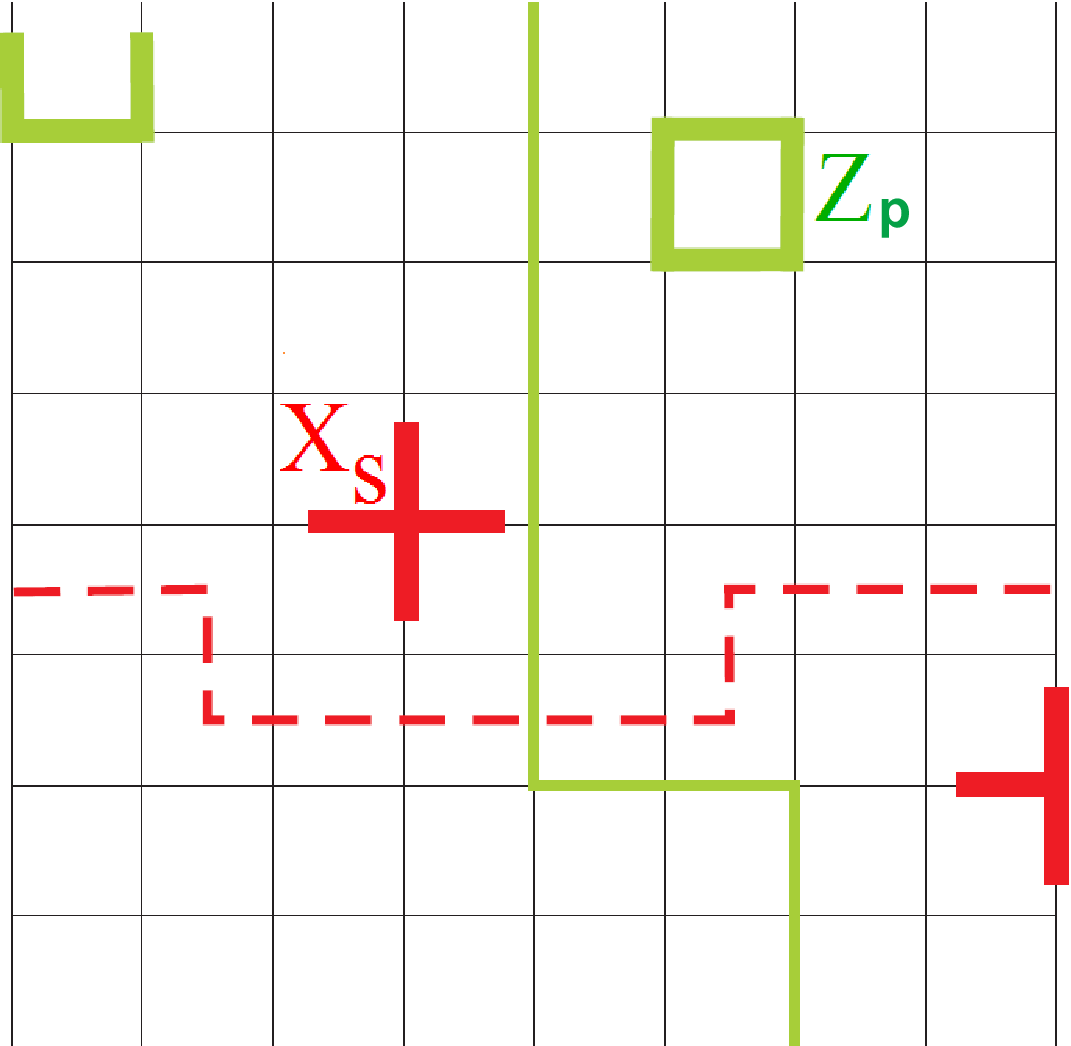}
  \caption{\label{fig:planar_code} Planar code. A code space is given by eigenvectors 
  of all star and plaquette operators with eigenvalues $+1$. 
  The lines represent exemplary logical $X$ (dashed line) and $Z$ (solid line) operators on the code space. 
  They are  given by homologically nontrivial loops, i.e. ones connecting opposite boundaries, in 
  the original and dual lattice, respectively. Construction of $X_{s}$, $Z_{p}$ observables is showed.}
\end{figure}

Here we present a much simpler scheme based on a concept of the topological code 
discovered by Kitaev \cite{KitaevFT} and developed in \cite{KitajewPreskill-FTmem}. 
In its version, called the planar code, qubits are situated on links of a 2D lattice 
(see Fig. \ref{fig:planar_code}). 
To maintain the encoded quantum information it is enough to measure 
repeatedly local four-qubit observables of two types - the plaquette observables $Z_p$ 
and the star observables $X_s$:  
\be
X_s=\ot_{l\in s } \sigma^x_{l},\quad
Z_p=\ot_{l\in p } \sigma^z_{l},
\ee
Here $s$ is associated with a vertex and it denotes all links that touch the vertex, while $p$ denotes all links that form the plaquette (Fig. \ref{fig:planar_code}).
In \cite{FowlerWHLMH2011-surface-comm} this scheme was applied to long distance communication, and it was analysed how the distance depends on the error rate. 
However the authors considered communication of logical qubits (i.e. encoded ones). To estimate the fidelity of communicating a physical qubit, or sharing 
entangled pairs, we need to complement this analysis with an encoding-decoding scheme. 
Such a scheme was proposed in \cite{KitajewPreskill-FTmem}. However it 
was relatively complicated in comparison with the simplicity of a planar code
and the scheme of maintaining the logical qubit. Here we shall propose an encoding-decoding scheme which is as close as possible to the simplicity of the Kitaev code. Namely, we shall use only {\it ideal} (i) measurement of 
$X_s$ and $Z_p$ operators, (ii) measurement and preparation of  $|0\>,|1\>$ and $|\pm\>$ states 
(where $|\pm\>=\frac{1}{\sqrt{2}}(|0\>\pm|1\>)$), (iii) bit and phase flips conditioned 
on measurement outcomes. The latter are needed in the one-shot scenario presented above only at the very last stage, where correction is applied to
the decoded qubit. However, it is not necessary if we want to use the scheme 
for quantum key distribution.\\

\begin{figure}
  \centering
  \includegraphics[scale=0.85]{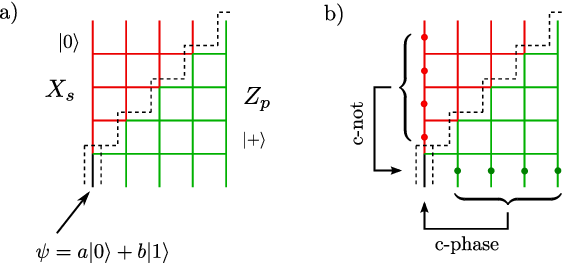}
\caption{\label{fig:diagonal-code} a) One shot encoding protocol. 
b) One shot decoding protocol. Red edges represent qubits originally prepared in $|0\rangle$ state. Green edges represent qubits originally prepared in $|+\rangle$ state. Black edge represents a physical qubit we want to encode/decode.}
\end{figure}

{\it One shot encoding.} The lattice is divided into three parts ( see Fig. \ref{fig:diagonal-code}): lower triangle (green qubits), upper triangle (red qubits),  and the qubit to be encoded (black one).

The procedure is as follows: (i) We measure all $X_s$ which include at least one red qubit, i.e. qubit from the upper triangle (there is no point in measuring $X_s$ within green region since the outcomes of such measurements are already known due to ideal state preparation).
(ii) We measure all $Z_p$ which include at least one green qubit, i.e. qubit from the lower triangle (the outcomes for red region are already known). (iii) We apply phase-flips to red qubits along arbitrarily chosen paths 
joining the $X$-defects. 
(iv) We apply bit-flips to green qubits along arbitrarily chosen paths 
joining the $Z$-defects.

{\it One shot decoding.}
We measure the lowest row of the qubits in $|\pm\>$ basis (except for the black qubit), 
compute parity, and flip phase of black qubit, when the parity is odd. 
Then we measure the leftmost column of the qubits in $|0\>,|1\>$ basis, 
compute parity, and flip the bit of the black qubit when the parity is odd. 

Below we prove that the above procedures correctly encode and decode a qubit in an unknown state, in a regime where the noise acts on the data qubits only.  

\begin{proposition}
The 'one shot encoding' procedure encodes the black qubit $\psi=a|0\> + b|1\>$ into a superposition of codewords  $a|0\>_L+ b|1\>_L$.
\label{prop:Kitaev_encoding}
\end{proposition}

{\bf Proof.}
Clearly after the above procedure we are in the code, as all defects are removed and 
therefore all stabilizers are set to $+1$. Then, it is enough to check that states $|0\>,|1\>,|+\>,|-\>$  are correctly encoded (cf. Lemma \ref{lem:fidelity} below). 
Let us first consider initial states $|0\>$ and $|1\>$. 
We  have to show that the value of a chosen logical operator $Z_L$ is $+1$ or $-1$, respectively. 
We can choose the operator along the leftmost vertical line,
which means that we need to check the bit parity of this line.
In the first stage some $X_s$ operators are measured, then phase flips are applied, 
and finally $Z_p$ operators are measured. (Note, that no bit-flips are applied to this line.)
The phase flips do not affect parity. 
From Lemma \ref{lem:x-meas} we conclude that the above measurements result in applications of $X_s$'s and $Z_p$'s to the code. 
Clearly, only an application of  $X_s$'s can affect bit values of the line. 
Since applied $X_s$ always touch two qubits from the line, they do not change the parity. 
Now, the initial parity is equal to the bit value of the black qubit. 
This proves that $|0\>$ and $|1\>$ are mapped into logical states $|0\>_L$ and $|1\>_L$
of the total code. 

The proof that $|+\>$ and $|-\>$ are correctly transferred is analogous and can be performed by examining the phase parity of the lower horizontal line (in a dual lattice). $\blacksquare$\\

\begin{proposition}
The 'one shot decoding' procedure decodes the superposition of codewords  $a|0\>_L+ b|1\>_L$
into the state  $\psi=a|0\> + b|1\>$ of the black qubit. 
\label{prop:Kitaev_decoding}
\end{proposition}

{\bf Proof.}
To this end we need to show that it 
sends $|0\>_L$ and $|1\>_L$ into $|0\>$ and $|1\>$, respectively. 
The proof that $|\pm\>_L$ is sent into $|\pm\>$ is the same. 
Again, by Lemma \ref{lem:fidelity}, having correctly transferred those four states, we obtain that all states  are also correctly transferred.
If the code is in logical state $|0\>_L$, then the left-most vertical line if measured 
would give an even number of $1$'s. Thus, if we measure all qubits from the line but the black one, the measured parity must be equal to the bit of the black qubit.
But we want to get $|0\>$, i.e. trivial parity. 
Hence, we have to apply the bit-flip operation, if the measured parity is nontrivial. 
The same reasoning works for initial logical state $|1\>_L$: The parity of the whole line is 
odd, thus if we want to have the bit value of the black qubit equal to 
that parity, we need to flip the black qubit, when the parity of other qubits is odd. $\blacksquare$\\

\begin{lemma}\cite{Hofmann2004-fidelity}
\label{lem:fidelity}
Consider a completely positive map $\Lambda$ on a single qubit, and define 
$F(\psi)=\<\psi|\Lambda(|\psi\>\<\psi|)|\psi\>$. Let $F_x,F_z$ be given by
\be
F_x=\frac12(F(|+\>)+F(|-\>))\quad
F_z=\frac12(F(|0\>)+F(|1\>)),
\ee
where $|\pm\>$ are eigenstates of $\sigma_x$ and $|0\>, |1\>$ are eigenstates of $\sigma_z$. 
Then:
\be
\overline F\geq F_x + F_z -1
\ee
where $\overline F=\int F(\psi) {\rm d}\psi $ is the fidelity averaged over uniformly chosen input states. 
\end{lemma}

\begin{lemma}
\label{lem:x-meas}
Let a system of $n$ qubits be in some bit basis state. 
Let $A$ be a subsystem. Then measuring $X_A\equiv \ot_{i\in A}\sigma_x^i$
leads to a superposition of two strings: the initial one, 
and the one flipped by $X_A$. 
\end{lemma}

{\bf Proof.} Let us write  $X_A=P_+-P_-$. Then we have $P_\pm=\frac12(I\pm X_A)$,
which proves the lemma. $\blacksquare$

\subsection{Teleportation description of encoding procedure}
Below we present a detailed description of the procedure of encoding a qubit in an unknown state into a planar Kitaev code as an equivalent to teleportation.
Let us first present teleportation in terms of the stabilizer formalism. 
To this end, recall that common eigenstates of operators $X_1X_2$ and $Z_1Z_2$ are Bell states \cite{Nielsen-Chuang}. Hence, teleportation can be viewed as follows.
Let qubit number 1 be the qubit to be teleported (held by Alice) 
and the qubits number 2 and 3 be the ones in maximally entangled state (qubit 2 owned by Alice and qubit 3 by Bob). Then teleportation is obtained by the following protocol:
First $X_1X_2$ is measured, and if the outcome is $-1$, a transformation $Z_2Z_3$ is performed.
Then $Z_1Z_2$ is measured, and, if the outcome is $-1$, a transformation $X_2X_3$ is performed.
These operations transform qubit 3 to initial state of qubit 1 and qubits 1 and 2 to an initial Bell state of qubits 2 and 3.

\begin{figure}[h]
  \centering
  \includegraphics[scale=0.8]{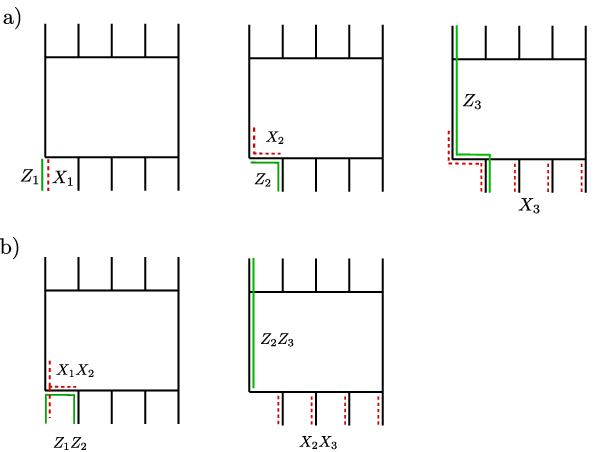}
  \caption{\label{fig:teleportation} Single shot encoding as teleportation.
  The black lattice symbolises the two-qubit code resulting from measuring all syndromes 
  but ones touching the physical qubit. a) Three qubits needed for teleportation; b) Two stages 
  of encoding can be interpreted as operations that perform teleportation.}
 \end{figure}

Now, let us consider our single shot encoding procedure. Note that we can first (i) Measure all the syndromes 
that  do not touch the physical qubit and remove defects. Then the remaining part of the encoding procedure consists of the following stages:
(ii) Measuring syndromes that touch physical qubit; (iii) Removing the obtained defects. 
More precisely,
we measure a single syndrome $Z_p$ which touches our physical qubit. If the syndrome 
is nontrivial, then we move it away by applying bit-flips to the lowest path  of the qubits (in the dual lattice). 
Then we measure a single syndrome $X_s$ which touches the physical qubit and, if the syndrome 
is nontrivial, we apply phase flips to the left-most vertical line of qubits (in the original lattice).

Let us now show that this is teleportation. To this end we have to determine three qubits. 
The first stage prepares a two-qubit code:
one qubit is described by logical operators $X_2$, $Z_2$ and the other by $X_3$,$Z_3$ 
(see Fig. \ref{fig:teleportation}). 
These  two qubits will be qubits number $2$ and $3$ and, because the code is a common eigenstate of operators $X_2X_3$ and $Z_2Z_3$, the qubits turn out to be in a maximally entangled state. 
Our physical qubit, associated with operators 
$X_1$, $Z_1$, is the one to be teleported (the operators defining the three qubits are visualised in 
Fig. \ref{fig:teleportation}a). Then, we notice that the operators $X_1X_2$ and $Z_1Z_2$ 
take exactly the form of syndromes $Z_p$ and $X_s$ touching the physical qubit, thus corresponding to the
stage (ii) above, while $X_2X_3$ and $Z_2Z_3$ are exactly those flipping bit and phase respectively, 
along the appropriate paths, as is done in the stage (iii). 

\section{ Fault tolerant encoding/decoding schemes for 2D Kitaev code.}\label{sec:4}
So far we have assumed that the operations (i)-(iii) defined in Section \ref{sec:2} are ideal. Here, we consider the case, where qubits are subjected to error (including preparation), and the measurement readout as well as performing corrections is non-ideal. We shall develop the original ideas of \cite{KitajewPreskill-FTmem}, the implementation 
of \cite{fowler-2008}, and obtain our main result, i.e. lower bound for storage total fidelity, given by Proposition \ref{prop6} in Section \ref{sec:d}. To this end we shall introduce a new metric on space-time lattice. The metric is designed to take into account effects of encoding and decoding stages, therefore it differs from standard, taxicab metric on initial and final time slices. After a brief overview of the protocol illustrated on the example of Kitaev code, we present the idea using the repetition code, and then apply it to the Kitaev code. We restrict ourselves to a scenario, where errors on the syndromes do not propagate back to the code. I.e., we imagine that a syndrome is measured in a noiseless way, and the classical  
outcome of the syndrome measurement is flipped with  some probability. However propagation issues
can be addressed as in \cite{fowler-2008}. 

\subsection{Algorithm: an overview}\label{sec:4a}
\begin{figure}[h!]
\includegraphics[scale=0.4,angle=0]{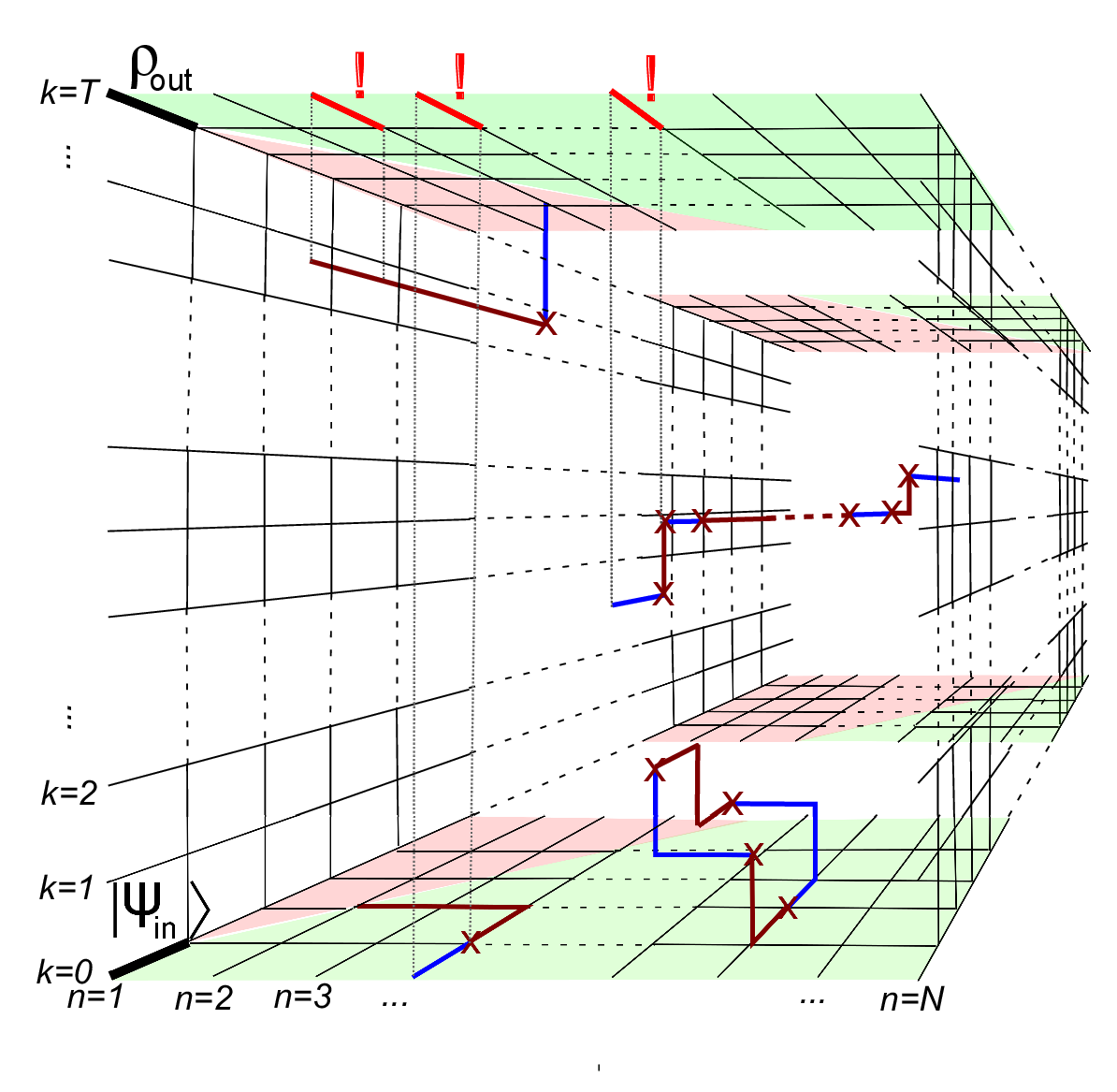}
\caption{\label{kit}
Structure of the algorithm of state $|\Psi_{in}\rangle$ protection through time $T$. Horizontal links represent qubits, vertical links store information about parity measurement outcomes performed on every time slice $k$ on every vertex ($X_{s}$) and plaquette ($Z_{p}$) (cf. Fig.\ref{od}).
At time $k=0$, encoding of a black qubit in state $|\Psi_{in}\rangle$ state into a system of $M$ qubits is performed by $X_{s}$ and $Z_{p}$ measurements on qubits prepared in $|+\rangle$ ($|0\rangle$) states, if a qubit belongs to a green (red) region. Defects (red crosses), marking ends of error chains $E$ (red lines), are determined if the measured parity is odd (in the picture, only exemplary defects and chains for phase protection part of the algorithm are depicted). For higher time steps, defects are determined whenever a parity measurement outcome changes from one slice to another. At time $T$, single-qubit measurements in $X$ basis (green region) and $Z$ basis (red region) are performed, and values of corresponding $X_{s}$ and $Z_{p}$ operators are calculated, determining defects on $T-th$ slice. The set of most probable error chains $E_{min}$ (blue lines) is determined classically by connecting defects with themselves or with a nearest boundary (which, in the case of phase protection, is a front or back \textit{rough} boundary, together with green regions of first and last time slices - as defects appear there with probability $\frac{1}{2}$  on every vertex (not marked in the picture)). $E_{min}$ is projected onto 'front row' ('left column') qubit line of the $T-th$ surface, leading to a  flipping (marked by exclamation marks) of a single-qubit parity outcome registered at time $T$ whenever the number of projections on a particular qubit is odd, and flipping the black qubit under the same condition. Parity on the 'front row' (left column) is calculated, and when it is odd, black qubit is flipped, which finalises decoding of state $\rho_{out}$ from a system of $M$ qubits. Non-trivial chains from $E+E_{min}$ are showed: short ones, connecting front \textit{rough} boundary with green regions of encoding/decoding slices, and one joining opposite \textit{rough} boundaries directly. Trivial (closed) chain is showed as well.}
\end{figure}
\begin{figure}[!]
\includegraphics[scale=0.4,angle=0]{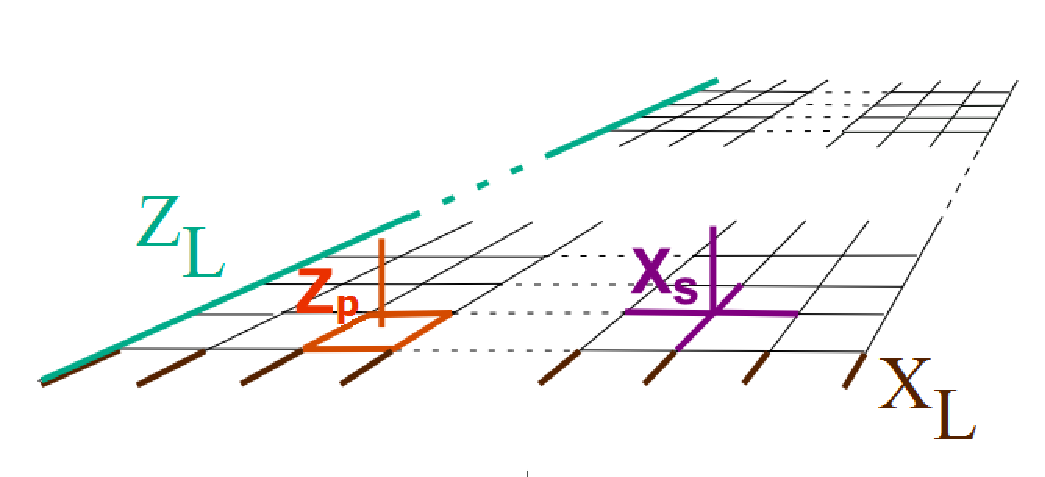}
\caption{\label{od}
Parity operators $X_{s}=\otimes_{i}X_{i}$ and $Z_{p}=\otimes_{i}Z_{i}$, defined on every time slice. Index $i$ labels qubits situated on links adjacent to a vertex (for $X_{s}$) and qubits forming boundary of a rectangle (for $Z_{p}$). 
Examples of logical operators defined in the codespace: $Z_{L}$ (green line), connecting front and back \textit{rough} boundaries, $X_{L}$ (brown line), connecting left and right \textit{smooth} boundaries.  
}
\end{figure}

\tblue{
We consider space-time structure for Kitaev code, where horizontal slices represent subsequent time steps (see Fig.\ref{kit} and Fig.\ref{od})}. There are vertical links: the ones connecting nodes represent bits where $X_s$ syndrome is collected, 
the  ones connecting plaquettes represent bits where $Z_p$ syndrome is collected. 
In the first and the last slice, we divide qubits into three parts, as in Fig. \ref{fig:diagonal-code}:
black qubit (the one to be encoded/decoded), red and green qubits.

\tblue{
In zero-th time step (lowest slice) green qubits are prepared in $|+\>$ states and red ones in $|0\>$ states. 
In between slices, syndromes $X_s$ and $Z_p$ are measured for {\it all} stars s and plaquettes p (unlike in the noiseless case, where it was sufficient to measure only that ones which include either red or green qubits). After the last slice, green qubits are measured in $|\pm\>$ basis 
and syndrome $X_s$ is computed, while red qubits are measured in $|0\>,|1\>$ basis and the values of $Z_p$ syndrome is computed. 
}

\tblue{
Now we will present the algorithm which corrects phase 
(the algorithm to correct bit is analogous). Let $S$ be the set of vertical links with nontrivial syndrome $X_s$. 
We have $\partial S=\partial E$, where $E$ is the set of links, where faulty syndrome was measured 
or a phase error for a qubit occurred. We select the set $E_{min}$, such that $\partial S=\partial E_{min}$.
Then we compute phase-parity of the front row of qubits based on the last measurement, 
and correct it by flipping, whenever projection of $E_{min}$ onto the first row of a single horizontal slice contains odd 
number of links.  Finally, we apply the phase flip to the black qubit, when the corrected parity is odd. 
}

\tblue{
The set $E_{min}$ is chosen in the following way. 
To $i$-th qubit we assign the weight $-\log \frac{p_i}{1-p_i}$ \cite{KitajewPreskill-FTmem}, where $p_{i}$ is the probability of an error on the qubit. For red qubits $p_{i}=1/2$, as they are prepared in $|0\>$ states. For the black qubit $p_{i}=p$ , as we assume it is subjected to the memory error $p$, and for the rest of the qubits $p_{i}=p$, as they are exposed to either preparation error $p$ (horizontal links) or syndrome measurement error $p$ (vertical links).
Now we choose $E_{min}$ in such a way, that it minimises the sum of weights (if there is more than one of the same weight 
we choose one of them at random).}

\tblue{
We now want to estimate the probability of the phase error of the black qubit under the described algorithm.
When is the phase wrongly decoded? This happens, if the set $E_{min}+ E$ (symmetric difference of sets $E$ and $E_{min}$)  
contains an odd number of links whose 'position' belongs to the first row, and 'time' is arbitrary. 
Conversely, we are sure that there is no error, if there are only homologically trivial loops in the set  $E_{min}+ E$,
since such loops cross the first row always even number of times.
So for sure the probability of error is no greater than probability of occurring of the set  $E_{min}+ E$
with a nontrivial loop. In Section \ref{sec:d} we prove the following bound for probability of such event in a limit of large code size}

\be
Prob(\text{nontrivial loop})\leq 2p+\frac{2\alpha^2(2-\alpha)}{(1-\alpha)^{3}},
\label{eq:boundphase}
\ee
where $\alpha=12\sqrt{p(1-p)}$.

As it should be, this probability scales linearly with the error rate. From (\ref{eq:boundphase}) it follows that for the probability of phase error \textit{Prob} to be less than $\frac{1}{2}$, the probability of error on single qubit $p$ needs to be bounded by the value $p=0.00042$. Let us quickly explain, how (\ref{eq:boundphase}) is obtained from the algorithm. In the fault tolerant storage as in Ref \cite{KitajewPreskill-FTmem}, 
we have probability of error tending to $0$, and this follows from the fact that 
the nontrivial loops are long, and their probability vanishes. 
In our case we have two types of paths: those that do not touch the first and the last time slice, 
and those that do touch either of them. Former paths appear with the probability proportional to $(\frac{p}{1-p})^L$, and this is like in the 
storage problem. However, latter paths have different probability for the part that lies in the first and in the last slice - their probability is proportional to 1 for slice regions where measurements are performed in complementary basis, so that the probability of such path scales like $(\frac{p}{1-p})^l$, where $l$ is the length of the part that does not belong to these regions. Thus the probability (\ref{eq:boundphase}) comes from the sum over all possible 
paths that start at the boundary and end up at the first or last slice region of measurements performed in complementary basis. There are very short paths there, including 
those of length $1$: one of them reflects the physical error that attacks the qubit in the first step,
when it is bare, and the syndrome is measured on it; the second is the dual path
that touches the last time slice and reflects the error which is acquired when other qubits are measured at the very end. The situation described above can be also explained from the following geometric point of view: The weights that we attribute to 
qubits constitute a sort of a metric, and the probability of appearing of a given path scales exponentially with
the length of the path in this metric. The nontrivial paths that do not touch green area of first or last slice 
are long, and therefore their probability asymptotically vanishes. The more they touch the green area, 
the shorter -  in this metric -  and therefore more probable they become, 
and formula \eqref{eq:boundphase} bounds the probability that such paths appear. 
In Section \ref{sec:c} encoding-decoding procedure proposed above will be discussed in more details.

\subsection{Analytical lower bound on state encoding fidelity: repetition code}\label{sec:b}

\begin{figure}[h!]
\includegraphics[scale=0.4]{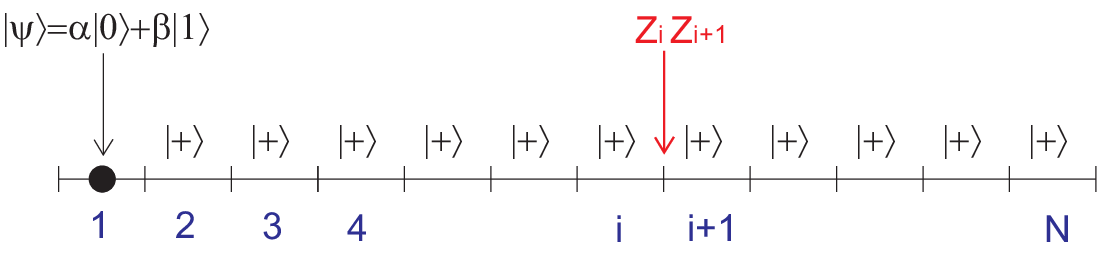}
\caption{\label{repetideal} The encoding procedure into a repetition code with an ideal syndrome measurement.
}
\end{figure}

In order to present a basic intuition laying behind our encoding-storage-decoding procedure for Kitaev code, below we present its analogue for repetition code. 

Suppose we want to encode a physical qubit in a state $|\Psi\rangle=a|0\rangle+b|1\rangle$ into a logical qubit $a|0\rangle|0\rangle...|0\rangle+b|1\rangle|1\rangle...|1\rangle$. Let us first assume that we measure syndromes with zero probability of error. We encode the physical qubit in the following way. On the right of the physical qubit, we prepare $N-1$ qubits in a state $|+\rangle...|+\rangle$. We denote the physical qubit by $1$, the next qubit by $2$ and so on. Then we measure operator $Z_iZ_{i+1}$ on each pair of neighbouring qubits. We can represent it graphically by drawing a line consisting of $N$ horizontal links and $N-1$ vertices, see Fig. \ref{repetideal}. Qubits correspond to the links, while syndrome measurements correspond to the vertices. Nontrivial error syndromes correspond to defects placed on the vertices. 

We are going to correct errors by optimally connecting all the defects. Optimal calculation of this set of connections must depend on error probability associated with each link. We thus perform a matching of defects by the shortest paths; length of a path is defined as a sum of weights of links that it is constituted of. Weight of a link is equal to $-\log \frac{p_i}{1-p_i}$ \cite{KitajewPreskill-FTmem}, where $p_i$ is probability of an error on every qubit. We assume that a physical qubit $1$ is subjected to a storage error $p$, i.e. the weight of the first link is equal to $-\log \frac{p}{1-p}$. On the other hand all error syndromes have random values, i.e. an error syndrome can have value $1$ with probability $\frac{1}{2}$ and value $-1$ with probability $\frac{1}{2}$. Hence, weights of other links are equal to zero. We correct errors by connecting all defects by the shortest path according to the metric given by weights, which is equivalent to moving all defects to the right. 

\begin{figure}[h!]
\includegraphics[scale=0.4]{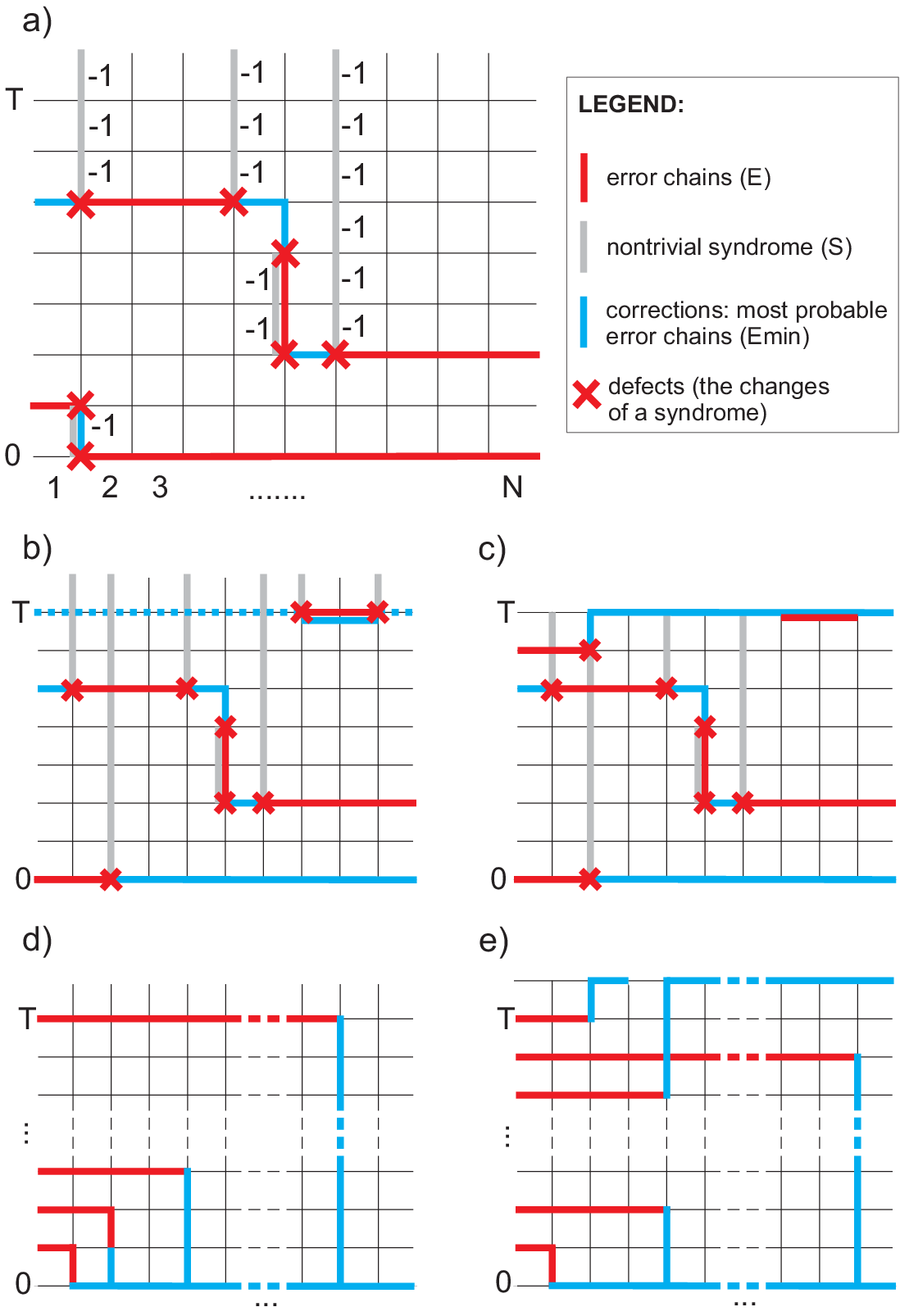}
\caption{\label{repet} a) Scheme for encoding a state $|\Psi_{in}\rangle$ of qubit placed on the bottom left corner into a repetition code. Qubits are represented by horizontal links, vertical links correspond to measurement outcomes of $Z_iZ_{i+1}$ operators acting on a pair of adjacent qubits. Red horizontal links represent errors that arise on qubits; red vertical links are associated with errors that occur on syndrome measurement; grey vertical links define places where nontrivial syndrome is measured; the changes of them are signed by red crosses. Blue lines represent chains of our corrections performed on the code. Both depicted non-trivial loops of errors lead to a failure of the encoding procedure. b) Single qubit measurements on the last time line are performed in $|0\rangle$, $|1\rangle$ basis. Using results of these measurements, perfect syndrome measurement outcomes are obtained. A boundary of a set $E$ is equal to syndrome changes marked by red crosses. Connecting defects by $E_{min}$ leads to either trivial (e.g. red line + solid blue line) or nontrivial (e.g. red line + blue dashed line) paths. c) Single qubit measurements on the last time line are performed in $|+\rangle$, $|-\rangle$ basis -- syndrome measurement outcomes are random there. Thus we can neglect them and assume that last syndrome measurement is performed on the ($T-1$)-th line. Some errors can cross the last time line without causing defects. d) Selected nontrivial paths in a case of ideal syndrome computation in the last time step, as described in panel (b). Bottom line (except of the first link on the left) is additionally treated as a boundary. Connecting defects by $E_{min}$ leads to paths that may have ends either on the left and right boundaries, or on the top boundary. e) Exemplary nontrivial paths in the case of lack of syndrome measurement on line $T$. There are two extra boundaries: 'upper' and 'bottom'.
}
\end{figure}

Let us now assume that we measure an error syndrome with some small, non-zero probability of error - for simplicity we take the probability of an erroneous measurement equal to the probability of an error on a physical qubit. In such a case we need to measure error syndrome many times. We can represent it graphically in the following way. We draw a lattice of horizontal links (associated with qubits) and vertical links (related to syndrome measurements in subsequent time steps $0,...,T$) - see Fig. \ref{repet}a. Horizontal links marked in red indicate selected qubits on which an error occurred, red vertical links correspond to erroneous syndrome measurements, and grey vertical links to places where measured parity is odd. Nontrivial error syndromes in the first step and changes of error syndromes in all subsequent steps correspond to defects placed on vertices (red crosses). The set of errors $E$ consists of all red links, and the set of non-trivial syndromes $S$ consists of grey links. We notice that the boundary of $E$ is defined by nodes where syndromes change, i.e. red crosses.

Having recorded the defects, we perform a correction by fliping qubits along chains from a set $E_{corr}$, which has the same boundary as the set $E$. As already announced, we can maximise the probability of successful recovery by calculating a set $E_{corr}$ as the set of most probable error chains, i.e. by taking $E_{corr}=E_{min}$. The elements of $E_{min}$ are chosen along the shortest paths according to the metric given by the following weights. Weight of the first link on the first horizontal line is equal to $-\log \frac{p}{1-p}$ and weights of the other links on the first horizontal line are equal to zero. After the first step probability of an error on qubits and probability of an erroneous measurement is equal to $p$. Hence, the weight of all other links is equal to $-\log \frac{p}{1-p}$. One can notice that in some regions of the lattice the ``visual'' lengths of the paths differ from the lengths (weights) determined by the metric, see the time slice $0$ 
in Fig. \ref{repet}b or the first and the last time slice in Fig. \ref{repet}c, where the paths of corrections are going to the right due 
to the metric (which identifies all the nodes but one) even though they are visually much longer than potential 
correction paths directed to the left.

Let us assume that the measurements of $Z_iZ_{i+1}$ in the last step are perfect. This can be achieved by classical (thus with no error) calculation of syndrome using outcomes of single qubit measurements performed in $Z$ eigenbasis: $|0\rangle$, $|1\rangle$ (with error $p$) on every qubit placed on the last line. In this case every error chain creates a defect or pair of defects. Hence a boundary of a set $E$ is equal to the set of defects, see Fig. \ref{repet}b. Therefore, we observe all chains in a set $E+E_{min}$ to be either closed or nontrivial paths. Since chains from $E+E_{min}$ have no boundary (because $\delta E = \delta E_{min} \Rightarrow \delta (E+E_{min})=0$), they cannot create a defect, and state of a system after this operation is in the codespace. 

An error on the logical qubit happens only if there is a nontrivial path consisting of an actual error ($E$) and the shortest paths by which we connected all defects ($E_{min}$). We have no more than $4^l$ nontrivial paths of length $l\geq 2$ going through the first vertex on the first horizontal line (the bottom nontrivial path in Fig. \ref{repet}d is an example of such a path), no more than $4^l$ nontrivial paths of length $l\geq 3$ going through the second vertex on the first horizontal line (see the second path from the bottom in Fig. \ref{repet}d), and so on. In general we have no more than $4^l$ nontrivial paths of length $l\geq k+1$ going through the $k$-th vertex on the first horizontal line. Additionally, we have no more than $T 4^l$ nontrivial paths of length $l \geq N$, where $T$ is a number of time steps (the upper nontrivial path in Fig. \ref{repet}a is an example of such path). There are no more nontrivial paths. Probability of the specified path of length $l$ is no greater than $2^l\sqrt{p(1-p)}^l$.  We can now bound the probability that there is nontrivial path by the following expression
\begin{eqnarray}
\label{eq:D1}
& \text{P}_{\text{SAP}}=\sum_{l=2}^\infty 4^l 2^l \sqrt{p(1-p)}^l \nonumber\\
& + \sum_{l=3}^\infty 4^l 2^l \sqrt{p(1-p)}^l +...+  \sum_{l=N}^\infty 4^l 2^l \sqrt{p(1-p)}^l \nonumber\\
& + T \sum_{l=N}^\infty 4^l 2^l \sqrt{p(1-p)}^l=\nonumber\\
& \frac{q^2}{1-q}+\frac{q^3}{1-q}+...+\frac{q^N}{1-q}+T\frac{q^N}{1-q}<\nonumber\\
& \frac{q^2}{(1-q)^2}+T\frac{q^N}{1-q},
\end{eqnarray}
where $q=8 \sqrt{p(1-p)}$.
This expression goes to $\frac{q^2}{(1-q)^2}$ for $p \ll 1$, $T=Polynomial(N)$ and $N \rightarrow \infty$. Hence, probability of an error on logical qubit scales linearly with $p$. It is worth to emphasise that eventually (in the limit of large code size N) only paths which start at the 'geometrical' boundary (on the left) and end at the 'bottom' boundary (introduced by randomness of syndrome measurements) contribute to the error, see Fig. \ref{repet}d.

On the other hand, if we consider a case when individual qubits on the last time line T are measured in $|\pm\rangle$ basis (which gives random outcomes as if the stabilizers were not measured at all), we can observe creation of a new type of paths in a set $E+E_{min}$, and as a result leaving the codespace, i.e. it can happen that an error chain is undetected at time $T$, and ends on the 'upper' boundary without creating a defect, see Fig. \ref{repet}c. In this case we can introduce 'bottom' boundary as well as 'upper' boundary arising due to randomness of measurements performed on qubits situated on first and last line. The bound for probability of error on logical qubit is now given by (\ref{eq:D1}) with additional factor $2$. The factor is due to the paths that join left boundary with the upper one (absent in the previous case) as depicted in Fig. \ref{repet}e. This case will be applied while describing encoding-decoding procedure in Kitaev code in Section \ref{sec:d}.

\subsection{Detailed description of fault tolerant encoding/storage/decoding algorithm}\label{sec:c}

Encoding and decoding protocol described briefly in Section \ref{sec:4a} aims at protecting a state of a chosen physical qubit $\rho$ from decoherence, modelled by a map $\Lambda(\rho)=p(1-p)X\rho X+p^{2}Y\rho Y+p(1-p)Z\rho Z+(1-p)^2\rho$ induced on a qubit in one time step; where $X,Y,Z$ denote Pauli matrices. The goal is achieved by encoding $\rho$ into a measurement-invariant, \textit{logical} subspace of a system of $M$ qubits, on which measurements are performed in order to detect bit errors (introduced by $X$) and phase errors (introduced by $Z$) resulting from the map $\Lambda$ acting on every qubit of the new system. Based on measurement outcomes, and under the assumption that the measurement itself is subjected to errors with probability $p$, we perform corrections in order to revert the action of the most probable errors on the logical subspace. In this way, after protecting the state for some time $T$, it will be possible to decode it from the $M$-qubit system into a chosen qubit.\\
  
{\it Code geometric structure}. In a particular geometry of a 2D planar code \cite{KitajewPreskill-FTmem}, qubits are represented by links forming horizontal slices of a code structure, visualised in Fig. \ref{kit}; a vertical axis $k$ describes time direction. For simplicity, we assume that the lattice is symmetric, i.e. there are $N$ qubits in every space direction (distance of a code = $N$), so $M=N^2 +(N-1)^2$. Joint phase- and bit- parity measurements are defined in the following way: for every node $s$ of a slice we measure an operator $X_{s}=\otimes_{i}X_{i}$, where index $i$ runs over every link (qubit) adjacent to a node $s$. Similarly, for every plaquette $p$ of a slice we measure an operator $Z_{p}=\otimes_{i}Z_{i}$, where index $i$ runs over every link (qubit) forming a plaquette boundary. This is shown in Fig. \ref{od}, which can be treated as a cut from Fig. \ref{kit}. Because $X$ and $Z$ anti-commute, measurements of $X_{s}$ ($Z_{p}$) reveal action of odd number of $Z$ ($X$) errors on qubits on which an operator $X_{s}$ ($Z_{p}$) is defined. Both joint phase - and bit- parity measurement operators act on $4$ qubits, with an exception of those defined by the nodes and plaquettes situated on the code boundaries: on the \textit{rough} boundary (front and back side of the code structure), $Z_p$ are formed by products of 3 one-site operators only; similarly, on the \textit{smooth} boundary (left and right side), $X_s$ are formed by products of 3 one-site operators. For every time step, measurements of $X_{s}$ as well as $Z_{p}$ (but in virtual lattice where $Z_{p}$ are associated with stars and $X_{s}$ with plaquettes) are performed on the slice, and results, called \textit{syndromes}, are stored in vertical links. By tracking the \textit{defects}, i.e. points where parity measurement outcomes change from one time step to another, we are able to identify boundary $\partial E$ of a set of error chains $E$. $E$ can be composed both of vertical and horizontal links of the code, as both qubits and parity measurement outcomes are subjected to error. In Fig. \ref{od} it can be seen that, in the geometry considered, $X$ ($Z$) error chains can start and end at smooth (rough) boundaries without causing any defects. 
Based on our knowledge about $\partial E$ and on probability of error on every link of the code, we can now classically determine a set of the most probable error chains $E_{min}$. Because all $X_{s}$ and $Z_{p}$ commute, this can be done independently for protection procedures against phase and bit errors. To determine $E_{min}$ it is thus sufficient to find the most probable set of links that connect all points from $\partial E^X$ with other points from $\partial E^X$ or rough boundary, as well as all points from $\partial E^Z$ with other points from $\partial E^Z$ or smooth boundary, where $\partial E^{Z(X)}$ refers to defects detected by $Z_{p}$ $(X_{s})$ parity operators.\\     

{\it Codespace protection}. The code (logical subspace of a system of M qubits) is described by logical bit ($X_{L}$) and phase ($Z_{L}$) operators. In the considered geometry, under an assumption that a state of the system is an eigenstate of all $X_{s}$ and $Z_{p}$ operators on a selected time slice (i.e. it belongs to a \textit{codespace}), these logical operators are defined by $X_{L}=\otimes_{i}X_{i}$ ($Z_{L}=\otimes_{i}Z_{i}$), where summation over index $i$ accounts for \textit{any} path within qubits of a chosen time slice and connecting opposite smooth (rough) boundaries; We are going to measure logical operators on lines presented in Fig. \ref{od}. $X_{L}$ and $Z_{L}$ commute with all $X_{s}$ and $Z_{p}$ stabilizers, which fulfils the requirement for the logical subspace to be measurement-invariant. The number of all $X_{s}$ and $Z_{p}$ operators on a chosen time slice is $M-1$ ($2^{M-1}$-dimensional subspace), which leaves room for exactly one qubit state to be stored in the logical subspace of the $2^M$-dimensional space of $M$ qubits. Because $X_{L}$ and $Z_{L}$ anti-commute (there is only one qubit that they act on jointly), as long as the requirement to be in the codespace is fulfilled, the parity of lines where $X_{L}$ ($Z_{L}$) is defined can be changed only by paths of $Z$ ($X$) operators that, projected onto the specific slice, form $Z_{L}$ ($X_{L}$) logical operators, i.e. chains that connect opposite boundaries (in a storage scenario, the probability of such chains to happen decreases to $0$ with growing code size $N$, under an assumption that single error probability $p$ is below some threshold value). We aim at achieving the requirement for a state of the $T$-th slice to be an eigenstate of all $X_{s}$ and $Z_{p}$ operators by connecting points from $\partial E^X$ with other points from $\partial E^X$ or rough boundary, and by connecting points from $\partial E^Z$ with other points from $\partial E^Z$ or smooth boundary, i.e. by flipping qubits along the most probable chains from a set $E_{min}$, which has the same boundary as $E$. After this operation, the set of errors is given by $E+E_{min}$, which represents a disjoint union of sets. For now, let us assume that all measurements at $T$-th slice are perfect, that is all error boundaries in the history $k\leq T$ are well defined (cf. Section \ref{sec:b}). Thus all chains in $E+E_{min}$ form paths (either closed or nontrivial). If the pairing of defects is fully successful in a sense that all paths in a set $E+E_{min}$ are closed, there will be no logical error, as closed paths on the code structure, projected onto the $T$-th slice, remain closed. Closed paths of single-link operators $Z$ ($X$) cannot influence a parity of $X_{L}$ ($Z_{L}$) since they intersect the selected line of logical operator an even number of times. However, if the pairing is not fully successful (there are \textit{non-trivial} paths in $E+E_{min}$ - those connecting opposite rough or smooth boundaries), the parity of the logical operator is changed; a logical error may occur. 
Remembering that the final goal of a protocol would be to recover, say at time $T$, a logical state of a single, physical qubit, we can lift the unrealistic assumption about the perfect quantum $X_{s}$ and $Z_{p}$ measurements on the last slice by classically (and thus with no error) calculating its outcomes from single qubit measurements at time $T$. Of course, these measurements destroy correlations between qubits, however, after decoding we no longer need to demand state of a system to be in the codespace. We define a single qubit measurement pattern so that at least on one line that $X_{L}$ ($Z_{L}$) operators are defined, i.e. no $Z$ ($X$) single qubit operators are measured. It provides that $X_{L}$ ($Z_{L}$) defined on this line commutes with the measurement and, in a way described by decoding procedure, its parity can be mapped into a chosen, physical qubit. In other words, decoding procedure does not map a parity of a really existing state from the codespace. It extracts information about the parity of $X_{L}$ and $Z_{L}$ that \textit{would have} characterised a codespace state if at time $k=T$ we had performed standard $X_{s}$ and $Z_{p}$ measurements, knowing which of the outcomes are correct and using that information together with registered $\partial E$ to apply corrections $E_{min}$ to the code. Of course, these imaginary  $X_{s}$ and $Z_{p}$ measurements on the last time slice should lead to the same error pattern as one caused by actual single qubit measurements.\\

{\it Description of the protocol}. Black qubit prepared in a state $|\Psi_{in}\rangle$ at time $k=0$ is situated on the front left corner of the plane (see Fig. \ref{kit}). Other qubits are prepared in $|+\rangle$ states ($X|+\rangle$=$|+\rangle$) (green region) or $|0\rangle$ states ($Z|0\rangle$=$|0\rangle$) (red region). Parity operators $X_{s}$ and $Z_{p}$ are measured; If the parity is odd, defect is recorded. 
For time steps $k = 1,. . . , T-1$, we measure parity operators $X_s$ and $Z_p$ for all stars $s$ and plaquettes $p$ and record defects (when parity measurement outcome changes from one time step to another). 
For time step $k=T$, we perform single qubit measurements in $X$ eigenbasis (green region) and $Z$ eigenbasis (red region). From outcomes of these measurements we calculate values of $X_{s}$ and $Z_{p}$ and record defects (if parity cannot be determined due to the presence of an element measured in different basis, we randomly choose between even and odd parity with equal probability. The same applies to a parity operator defined on the black qubit, which, as the only one at the $T$-th slice, is not measured). 

Having recorded the defects (which is equivalent to knowing $\partial E$), we calculate the set $E_{min}$ of most probable error paths that could have caused it. The optimisation is achieved by minimising the sum of the weights $-\sum_{i}\log{\frac{p_{i}}{1-p_{i}}}$ by a proper choice of paths of links connecting all points from $\partial E^X$ with other points from $\partial E^X$ or rough boundary, as well as all points from $\partial E^Z$ with other points from $\partial E^Z$ or smooth boundary ($p_i$ stands here for probability of error occurrence on link $i$). 
We assume that state preparation, storage and measurement are subjected to an error with probability $p$. Thus, the perfect minimum weight matching algorithm \cite{minimal_algorithm}, which determines $E_{min}$, is going to use link weights in the units of $-\log{\frac{p}{1-p}}$. Hence all links except from those located on the lowest and the highest slices have weight $1$. In a case of $k=0$ and $k=T$ slices, the algorithm used to construct $E_{min}^{X}$ ($E_{min}^{Z}$) assumes $0$ weights for all links from red (green) regions, as measurement of a state in a complementary basis gives random outcome ($p_{i}=\frac{1}{2}$). Thus, as $X$ ($Z$) error chains ending in the intersection between green and red region can be extended towards the rough (smooth) boundary with no cost (weight $0$), this effectively moves code boundaries across triangle-shaped regions of the highest and the lowest slices. To the green (red) links, for $E_{min}^{X}$ ($E_{min}^{Z}$) procedure, we attribute weight value $1$, as state preparation and measurement are faulty with probability $p$. We assume that the black qubit is subjected to a storage error $p$ during first and last stage of the protocol, so weight values $1$ are associated with it. Therefore, the metric used for perfect matching varies from taxicab metric only on the first and the last slice of the lattice, where presence of links with  weight 0 stems from encoding and decoding procedures, respectively. 
After calculating $E_{min}^{X}$ ($E_{min}^{Z}$), we make its projection onto the $T$-th slice 'front row' ('left column') qubit line. If the number of corrections projected into a particular qubit is odd, we flip its measurement outcome that was obtained at $k=T$ time (if the projections are done onto the unmeasured black qubit, we apply $X$ ($Z$) on it whenever number of projections from $\delta E^{X}$ ($\delta E^{Z}$) is odd). Neglecting the black qubit and basing on the modified single-qubit measurements at 'front row' ('left column') line on $T$-th slice, we calculate the parity of the line. If it is odd, we apply a $Z$ ($X$) operator on a black qubit, obtaining $\rho_{out}$.      
  
Before going into details of a rigorous proof of the $p$-dependent lower bound for probability of $\langle\Psi_{in}|\rho_{out}|\Psi_{in}\rangle= 1$ (success of the procedure), let us present the intuition behind the protocol. The aim of the $X_{s}$ and $Z_{p}$ parity measurements at time $k=0$ is to map the qubit state $|\Psi_{in}\rangle$ into the code (that would have existed at time $T$ had it not been destroyed by single qubit measurements). Preparation of 'left column' states $|0\rangle$ and 'front row' states $|+\rangle$ sets the dependence of the logical state of the code on the state $|\Psi_{in}\rangle$. Indeed, the black qubit and the 'front row' ('left column') qubits constitute sets of qubits on which the logical $Z_{L}$ ($X_{L}$) operator is defined, and we see that parities of those operators depend completely (in the case of ideal measurements) on the state $|\Psi_{in}\rangle$.      
In a fault-tolerant scenario we prepare half of the $k=0$ plane in $|+\rangle$ ($|0\rangle$) states in order to reduce probability that a chain from $E^{X}+E_{min}^{X}$ ($E^{Z}+E_{min}^{Z}$) crosses the 'front row' ('left column') line, changing in uncontrollable way its parity. This results in enlarging a distance between the line and the boundary (constituted for $E_{min}^{X}$ ($E_{min}^{Z}$) algorithm by green (red) area, cf. Fig. \ref{kit}) with growing $n$, enabling upper bound on probability of failure not to scale with the size of the code $N$. However, as short chains from $E+E_{min}$, that lead to logical error, are always present, probability of failure is non-zero. From a geometric point of view, the decoding procedure at $k=T$ is symmetric to encoding at $k=0$. The presented encoding procedure maps the parity of the hypothetical state from the codespace onto the black qubit, while destroying the coherences between the qubits. This can be seen in the following way: If the parity of a logical operator, characterising a state in the codespace, is originally odd (even), the procedure leads to changing black qubit parity from even to odd (odd to even), thus synchronising it with logical operator parity; it leaves it untouched when its parity is already synchronised. It should be noted that both encoding and decoding procedures do not discriminate between eigenstates of $X$, as well as between eigenstates of $Z$, thus enabling the coherence in the state $|\Psi_{in}\rangle$ to be sustained.     

Fig. {\ref{kit}} shows short chains from $E^{X}+E_{min}^{X}$ connecting the 'front row' line of qubits with the opposite rough boundary through the green areas of the $k=0$ and $k=T$ slice. Non-trivial loops of this kind are the most probable (i.e. the shortest among all non-trivial loops) for times $t<\frac{N}{2}$ and $t>\frac{N}{2}$, respectively. However, for intermediate times, the shortest non-trivial loops connect rough boundaries directly, without going through the green regions, as it is shown in Fig. {\ref{kit}}. The closed path is depicted as well in order to envisage that, when projected into $T$-th slice, it cannot influence the parity of any line that logical operator is defined at (it always acts on it even number of times).
  
As fidelity of a quantum process relies on measurement outputs of only two complementary sets of input states (cf. Lemma  \ref{lem:fidelity}), in order to prove the correctness of procedures for encoding, storage and decoding of eigenstates of $X$ and $Z$ operators it suffices to show the correctness of the algorithm for an arbitrary state $|\Psi_{in}\rangle$.\\

\subsection{Calculation of lower bound on the protocol fidelity.}\label{sec:d}

 \begin{proposition}\label{prop6}
An unknown quantum state is encoded from a single qubit into Kitaev 2D code of size $N$, stored through time $T$, and decoded into a qubit. With encoding/decoding realised by the algorithm described in Sec.\ref{sec:4a}, under the assumption of local Markovian noise in form $\Lambda(\rho)=p(1-p)X\rho X+p^{2}Y\rho Y+p(1-p)Z\rho Z+(1-p)^2\rho$, acting independently on every qubit in a single time step of storage, and state preparation and classical measurement error $p$, the fidelity $\overline{F}$ of the encoding/storage/decoding procedure satisfies    
 \begin{equation}\label{F}
\overline{F}\geq 1- 6p - \frac{2\alpha^{2}(5-3\alpha)}{(1-\alpha)^{3}}-2NT\frac{\alpha^{N}}{1-\alpha}-\frac{2\alpha^{N}(3\alpha-2)}{(1-\alpha)^{3}},
 \end{equation}
 with $\alpha=12\sqrt{p(1-p)}$.
 \end{proposition}

 {\bf Proof.}\\
To prove the above proposition, we will exploit Lemma \ref{lem:fidelity} by calculating the fidelity of encoding/storage/decoding the $|0\rangle$ and $|+\rangle$ states.\\

As it was announced, the part of the scheme protecting from phase flips relies on parity measurement of 'front row' qubit line at time $T$ (see Fig. \ref{kit}). We have to take into account all nontrivial paths in the set $E^{X}+E_{min}^{X}$ that result both from the interaction with environment and from the applied algorithm of calculating $E_{min}^{X}$ . Nontrivial paths, i.e. those having nontrivial projection on the 'front row' line at $k=T$, can start from any point $(n,k)$ on the front rough boundary and connect it either with the back rough boundary directly or with additional boundaries - red regions in $k=0$ and $k=T$ slices. We should sum over all possible ways $\eta_{n,k}^{l}$ of realising paths of length $l$ sufficient to reach another boundary, and take into account probability $prob(l)$ of occurrence of a path of length $l$ from the set $E^{X}+E_{min}^{X}$. Thus, probability of failure (phase flip) in encoding/storage/decoding the $|+\rangle$ state is bounded by
\begin{equation}
P_{fail}^{x}\leq\sum_{n=1}^{N}\sum_{k=0}^{T}\sum_{l=min(N,n+k,n+(T-k))}^{\infty}\eta_{n,k}^{l}prob(l),
\end{equation}
where $l$ stands for a minimal length of nontrivial loop which starts at a point $n,k$. From \cite{KitajewPreskill-FTmem} we conclude that the latter is bounded by $prob(l)\leq(2\sqrt{p(1-p)})^{l}$. As paths are realised in $3D$ structure, we overestimate $\eta_{n,k}^{l}\leq 6^{l}$. Only in a case of a single link error chain we take exact values $prob(l=1)=p$, $\eta_{n=1,k=0}^{l=1}=1$, $\eta_{n=1,k=T}^{l=1}=1$. Thus we obtain
\begin{widetext}
\begin{equation}\label{przeszacowanie}
P_{fail}^{x}\leq 2p+\sum_{k=1}^{T-1}\sum_{l=\min(N,1+k,1+(T-k))}^{\infty}\alpha^{l}+\sum_{n=2}^{N}\sum_{k=0}^{T}\sum_{l=\min(n,n+k,n+(T-k))}^{\infty}\alpha^{l}, 
\end{equation}
where $\alpha=12\sqrt{p(1-p)}$. Under the assumption $\frac{T}{2}>N-2$, RHS of (\ref{przeszacowanie}) can be expanded as 
\begin{eqnarray}\label{roww}
\begin{matrix}
2p+& 2\big([\alpha^{2}+\alpha^{3}+...]   & +[\alpha^{3}+...]+&...&  +[\alpha^{N-1}+\alpha^{N}+...]\big)&+(T+3-2N)(\alpha^{N}+\alpha^{N+1}+...)+\\
&+2\big([\alpha^{2}+\alpha^{3}+...] & +[\alpha^{3}+...]+&...&  +[\alpha^{N-1}+\alpha^{N}+...]\big)&+(T+5-2N)(\alpha^{N}+\alpha^{N+1}+...)+&& \\
&&+2\big([\alpha^{3}+...]+&...&  +[\alpha^{N-1}+\alpha^{N}+...]\big)&+(T+7-2N)(\alpha^{N}+\alpha^{N+1}+...)+\\
&&&&&+ ...+ \\
&&&&&+(T+1)(\alpha^{N}+\alpha^{N+1}+...)=
\end{matrix}
\end{eqnarray}

\begin{equation}
\begin{aligned}
&=2p+2\big(\underbrace{2\alpha^{2}+5\alpha^{3}+...+\frac{(N-1)N-2}{2}\alpha^{N-1}}_{\sum_{j=2}^{N-1}\frac{j(j+1)-2}{2}\alpha^{j}}\big)+\Big(N(T+1)-2\Big)(\alpha^{N}+\alpha^{N+1}+...+\alpha^{\infty})= \nonumber\\
&=2p+2\sum_{j=2}^{N-2}\frac{j(j+1)-2}{2}\alpha^{j}+\Big(N(T+1)-2\Big)\sum_{j=N}^{\infty}\alpha^{j}=\nonumber\\
&=2p+\frac{2\alpha^2(2-\alpha)+\alpha^{N}\big(N^{2}[1-\alpha]^{2}+N[1-\alpha^{2}]-2\alpha^2+6\alpha-2\big)}{(1-\alpha)^{3}}+\Big(N(T+1)-2\Big)\frac{\alpha^{N}}{1-\alpha}=\nonumber\\
&=2p + \frac{2\alpha^{2}(2-\alpha)}{(1-\alpha)^{3}}+NT\frac{\alpha^{N}}{1-\alpha}+\frac{\alpha^{N}\big(N^{2}(1-\alpha)^{2}+2N(1-\alpha)-4\alpha^{2}+10\alpha-4\big)}{(1-\alpha^{3})}.
\end{aligned}
\end{equation}
\end{widetext}
 
Now, one can regard calculating failure probability for $\frac{T}{2}\leq N-2$ as ignoring some layers of qubits from the front rough boundary of the code (cf. Fig. \ref{kit}) that are connected by the shortest error paths with the opposite rough boundary, rather than with red boundaries of lower (k=0) or upper (k=T) layers. This is equivalent to ignoring some terms of the form $(\alpha^{N}+\alpha^{N+1}+\dots)$ in (\ref{roww}). Therefore, RHS of (\ref{przeszacowanie}) in the case  $\frac{T}{2}\leq N-2$ is even smaller than (\ref{roww}). In the limit of large code size $N\rightarrow \infty$ we obtain RHS of (\ref{eq:boundphase}).\\

Analogously, in the case of the $|0\rangle$ state encoding/storage/decoding, the probability of failure (bit flip) is given by
\begin{eqnarray}
P_{fail}^{z}\leq 4p + 2\sum_{k=1}^{T-1}\sum_{l=min(N,1+k,1+(T-k))}\alpha^{l}+\nonumber\\
+\sum_{m=3}^{N}\sum_{k=0}^{T}\sum_{l=\min(N,m-1+k,m-1+(T-k))}\alpha^{l}=\nonumber\\
=4p + \frac{2\alpha^{2}(3-2\alpha)}{(1-\alpha)^{3}}+NT\frac{\alpha^{N}}{1-\alpha}+\nonumber\\+
\frac{\alpha^{N}\big(-N^{2}(1-\alpha)^{2}-2N(1-\alpha)+2\alpha^{2}-4\alpha\big)}{(1-\alpha^{3})},
\end{eqnarray}
where now all nontrivial paths start at points indicated by $(m,k)$.

From Lemma \ref{lem:fidelity} we have
\begin{eqnarray}
\overline{F}\geq (1-P_{fail}^{x})+(1-P_{fail}^{z})-1, 
\end{eqnarray}
which gives (\ref{F}).
 $\blacksquare$\\
 
We note two implications of the above proposition.\\

{\bf Implication 1}\\

\textit{
For $p\lessapprox 0.007$, $\alpha$ is smaller than $1$, and in the limit of large code size for $T$ polynomial in code size we have
\begin{equation}\label{bound}
\lim_{N\rightarrow \infty} \overline{F} \geq 1- 6p- \frac{2\alpha^{2}(5-3\alpha)}{(1-\alpha)^{3}},
\end{equation}
which yields $\lim_{N\rightarrow \infty} \overline{F}\geq 1- 1448p+\mathcal{O}(p^{\frac{3}{2}})$.}\\

It shows that the probability of failure is proportional to the probability of the shortest chain capable of creating logical error.
Let us note that RHS of (\ref{bound}) was largely underestimated and increases above $\frac{1}{2}$ for $p\leq 0.00021$.\\\\

{\bf Implication 2}\\

\textit{Size $N$ of 2D Kitaev code scales at most logarithmically with storage time $T$.}\\

If we manipulate size N and storage time T: $N^{'}=f_{N}N$, $T^{'}=f_{T}T$, 
in order to keep the lower bound (\ref{F}) constant we need, for $T\gg N$,  
\begin{eqnarray}
NT\alpha^{N}=N^{'}T^{'}\alpha^{N^{'}}, 
\end{eqnarray}
which implies
\begin{eqnarray}
0=\log_{\alpha}f_{N}+\log_{\alpha}f_{T}+N(f_{N}-1).
\end{eqnarray}
For $\alpha<\frac{1}{e}^{\frac{1}{\beta e}}$ we have $\log_{\alpha}f_{N}>-\beta f_{N}$ and  
\begin{eqnarray}\label{log}
f_{N}<\frac{-\log_{\alpha}f_{T}+N}{N-\beta}, 
\end{eqnarray}
for $\beta<N$. One can manipulate $\beta<N$ to apply the reasoning up to $\alpha\simeq 1\ \Rightarrow p\simeq 0.007$.

\section{Implications for entanglement percolation.}\label{sec:5}
In entanglement percolation \cite{AcinCiracL-perc} the idea is 
that in each node one performs operation consisting of constant number of 
elementary operations, i.e. it does not depend on 
the size of the network. The task is then to share entanglement 
between two distant nodes with the fidelity $F$, 
which does not decay when the network is enlarged. 

\begin{figure}
\centering
\includegraphics[width=8cm]{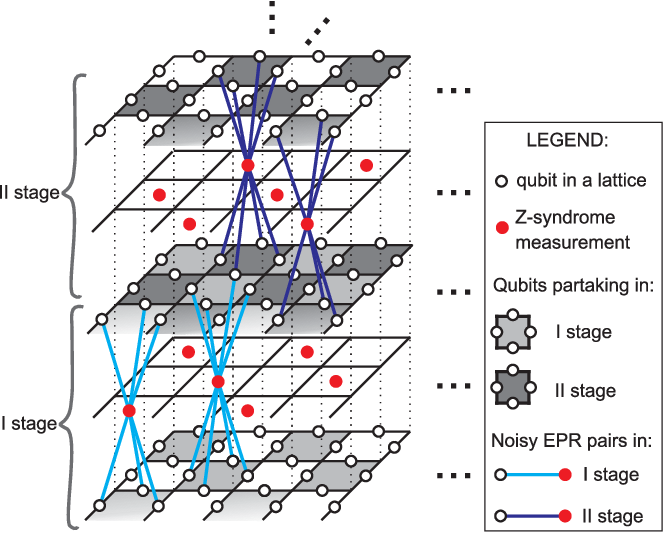}
  \caption{\label{fig:syndrom}Z-syndrome measurements via noisy EPR pairs.}
\end{figure}

According to \cite{perseguers-2008-78}, if we have computation scheme of a dimension $d$, which allows to preserve a qubit in time by means of local gates, then we can translate  it into a network of dimension $d+1$ that allows to share entanglement between two nodes of network. 
Our results now imply, that in the 2D EPR-network a node-to-node entanglement percolation 
is possible. In other words, if in the 2D square network the neighboring nodes
share an EPR pair, two nodes can share an entangled pair with the fidelity 
that scales reasonably with the error rate.

Since our scheme based on 1D architecture for fault-tolerant 
quantum computing uses concatenated codes, it is quite complicated. 
We have therefore also proposed a scheme of encoding/storing/decoding  qubit in 2 dimensions  by measuring local syndromes. Via the mentioned result of \cite{perseguers-2008-78} we obtain a scheme of distant communication in 3D EPR network. 
To see how the implementation of the idea of  \cite{perseguers-2008-78} would look in our case,
let us  describe a process of measuring $Z$ syndromes (see Fig. \ref{fig:syndrom}). The total 3D network would consist of 2D slices:
the code slices (white qubits in Fig. \ref{fig:syndrom}) interlaced with syndrome collecting slices (red qubits).
In first slice we consider half of the plaquettes  (so that they do not share qubits), and each plaquette from this set
is teleported to a single node in next slice. In that slice, the four qubits are jointly measured (i.e. syndromes are measured)
and they are teleported to four qubits occupying separated nodes in the next slice.  Then we do the same with second half of plaquettes.
So using 5 slices of network, with at most four qubit in the nodes, and with at most 8 EPR pairs connecting to a single node,
we are able to implement measurement of Z syndromes of 2D network.  Thus such a 3D network can simulate time evolution of 2D network, 
and the fidelity of the used EPR pairs is related to error rate on the 2D network.
Therefore our result on encoding/decoding implies  that 3D network with noisy EPR pairs allows communication over arbitrary distance. 

Let us stress that we do not need long-time quantum memory - instead we demand probability of a single error per time step to be below some threshold value. If this condition is satisfied, we only require local, short-time quantum memory. Our percolation scheme does not take into account effects of finite time of classical calculations; however in some applications (e.g. cryptography tasks) this is not the issue as corrections can be applied to classical values after the end of a protocol.

The communication scheme for encoded state was analysed in \cite{FowlerWHLMH2011-surface-comm}. 
Here we have complemented it  with a simple scheme of encoding/decoding an unknown state,
which allows to obtain node-to-node entanglement percolation. 
Our scheme, similarly to that of \cite{perseguers2009-fid}, 
needs three dimensions.  Two of them scale only logarithmically with the third one -- the 
distance between the nodes which want to share entanglement. If we change time into a space dimension, the logarithmic scaling is visible from the bound on $f_{N}$ in (\ref{log}). Our protocol is  more uniform than the one presented in \cite{perseguers-2008-78}, because the only operations are syndrome measurements for encoding, Pauli measurements for decoding and entanglement swapping. Indeed, we do not need to 
perform the flips in the encoding scheme: it is enough to store this information classically. Communication scheme of \cite{perseguers2009-fid} uses correlations between the nodes emerging from their initial preparation in a cluster state, whereas our method relies on EPR pairs shared by adjacent nodes.

\begin{figure}
\centering
\includegraphics[width=4cm]{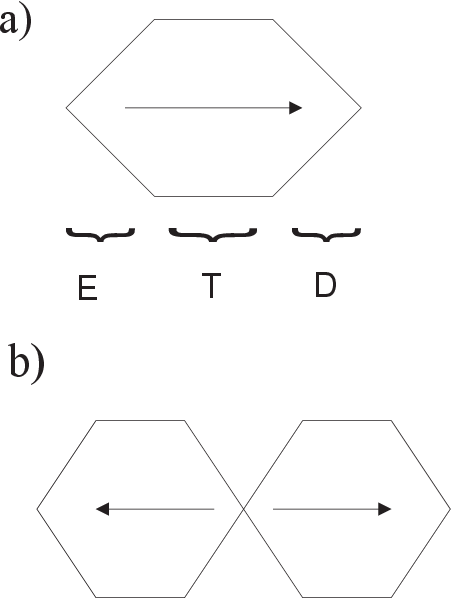}
  \caption{\label{fig:q-e} Transmitting qubits vs sharing ebits. By E,D, T we denote encoding, decoding and transmitting.}
\end{figure}

In order to transmit a qubit, it is enough to change time into one more space 
direction as in \cite{perseguers-2008-78}. To share entanglement, 
we need to propagate two qubits in opposite time directions. If we translate it into 
space, we obtain the following scheme: In one node an EPR pair is prepared 
and each of two qubits is transmitted towards nodes between which we want to share entanglement (see Fig. \ref{fig:q-e}).

\begin{figure}
 \centering
\includegraphics[width=5cm]{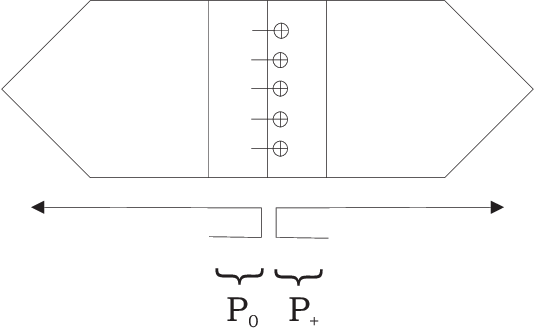}
\caption{\label{fig:no-encoding} Sharing e-bits without unknown-state encoding 
stage. Here $P_{0,(+)}$ mean the stage of preparing logical state $|0\>_L$ ($|+\>_L$).}
\end{figure}

There is an alternative scheme of sharing entanglement, 
which does not need encoding, but decoding only (see Fig. \ref{fig:no-encoding}). 
The idea is the following: We shall not start from a qubit which we then want to transmit. 
Rather, we shall prepare known encoded states. 
On one 2D plane we prepare $|0\>_L$ state (which in percolation picture will use 
up 3D network). This amounts to prepare all qubits in the state $|0\>$,  measure 
repeatedly star and plaquette observables and write down the syndrome.
On another 2D plane we shall prepare  
in an analogous way $|+\>_L$ state. Then C-NOTs will be applied bit-wise between the 
planes, so that we shall obtain the EPR encoded state between two logical 
qubits. Then we transmit the encoded qubits in opposite directions, and 
then localise them into two single qubits by our decoding method. The same of course can be done in the case of 2D network.

Finally, one should be aware that the scheme of quantum communication 
over networks inherits the problems of applicability of fault tolerant schemes to the 
Hamiltonian description of interaction of the system with environment, 
see e.g. \cite{AHHH2001,Alicki2004-fails,TerhalB-FT,Alicki-TB,AlickiLZ2005,AharonovKP-FT-NM} or a recent discussion \cite{FT_perpetual_motion_blog}.\\

\section{Conclusions}\label{sec:6} We have completed previous results on quantum communication by use of a 2D network without long-time quantum memory by
explicitly evaluating the encoding fidelity. We have also provided an elegant scheme of encoding/decoding 
 an unknown qubit into the 2D Kitaev code by measuring stabilizer operators, thereby providing a simple scheme for long distance communication in 3D. Our encoding scheme can also be used in quantum computing architecture based on 2D Kitaev codes, where encoding a proper superposition (followed by distillation) allows to implement a non-Clifford gate needed for universal quantum computing \cite{Bravyi}.\\

We would like to thank Robert Alicki, Stuart Broadfoot, David DiVincenzo, Ryszard Horodecki and Andrew Steane for helpful discussions.  This paper was supported by European Q-ESSENCE (248095) and by the Polish Ministry of Science 
and Higher Education Grant No. N202 231937. The final version of the paper was supported
by the Polish Ministry of Science and Higher Education Grant No. IdP2011 000361.
P. M. and A. P.  were supported by the Foundation for Polish Science International PhD Projects Programme co-financed by the EU European Regional Development Fund. Part of this work was done in National Quantum Information Centre of Gda\'nsk.

\bibliographystyle{apsrev}

\bibliography{long-crypto9-diagonal_2}

\end{document}